\documentclass{raa}            

\usepackage{listings}
\usepackage{graphicx,times}             
\usepackage{natbib}
\usepackage{amssymb,amsmath}
\bibpunct{(}{)}{;}{a}{}{,}

\usepackage{color}
\usepackage[pagebackref=true]{hyperref}
\usepackage{tikz}
\usetikzlibrary{positioning, chains, arrows.meta, shadows.blur, shapes.geometric, calc}

\definecolor{tianlaiBlue}{RGB}{41, 98, 155}
\definecolor{tianlaiBlueLight}{RGB}{91, 155, 213}
\definecolor{tianlaiGreen}{RGB}{76, 153, 0}
\definecolor{tianlaiGreenLight}{RGB}{169, 208, 100}
\definecolor{textDark}{RGB}{50, 50, 50}
\definecolor{arrowBlue}{RGB}{60, 120, 180}
\definecolor{arrowGreen}{RGB}{60, 140, 60}

\definecolor{ayopurple}{rgb}{0.5,0,0.5}

\begin{document}

  \title{21 cm Power Spectrum Analysis of North Celestial Pole Observations with the Tianlai Dish Pathfinder Array}

   \volnopage{Vol.0 (202x) No.0, 000--000}      
   \setcounter{page}{1}          

   \author{Guangzhi He 
      \inst{1,2}
   \and Shifan Zuo
      \inst{1,2}
   \and Jixia Li
      \inst{1,2}
   \and Yichao Li
      \inst{3}
    \and Furen Deng
      \inst{1,2}
    \and Shijie Sun
      \inst{1,3}
    \and Reza Ansari
      \inst{4}
    \and Olivier Perdereau
      \inst{5}
    \and Peter Timbie
      \inst{6}
    \and Albert Stebbins
      \inst{7}
    \and Ayodeji Ibitoye
      \inst{8,9,10}
    \and Fengquan Wu
      \inst{1}
   \and Yougang Wang
      \inst{1,2,3}
   \and Xuelei Chen
      \inst{1,2}
}

   \institute{
   State Key Laboratory of Radio Astronomy and Technology, National Astronomical Observatories, CAS, A20 Datun Road, Chaoyang District, Beijing, 100101, P. R. China; {\it sfzuo@bao.ac.cn,\it wangyg@bao.ac.cn, \it xuelei@cosmology.bao.ac.cn}\\
        \and   
    School of Astronomy and Space Science, University of Chinese Academy of Sciences, Beijing 100049, P. R. China\\
        \and
    Key Laboratory of Cosmology and Astrophysics (Liaoning) \& College of Sciences, Northeastern University, Shenyang 110819, China\\
    \and
    Université Paris-Saclay,  
Université Paris Cité, CEA, CNRS, AIM, 91191 Gif sur Yvette, France\\
    \and
    IJCLab, CNRS/IN2P3, Universit\'e Paris-Saclay, 
91405 Orsay, France\\
    \and
    Department of Physics, University of Wisconsin Madison, 1150 University Ave,Madison WI 53703, USA\\
    \and
    Fermi National Accelerator Laboratory ,P.O. Box 500, 
Batavia IL 60510, USA\\
   \and
   Department of Physics, Guangdong Technion—Israel Institute of Technology,
Shantou, Guangdong 515063, People’s Republic of China\\
   \and
   Centre for Space Research, North-West University, Potchefstroom 2520, South Africa \\
   \and
   Department of Physics and Electronics, Adekunle Ajasin University, P. M. B. 001, Akungba-Akoko, Ondo State, Nigeria\\
\vs\no
   {\small Received 2026 month day; accepted 202x month day}}

\abstract{The Tianlai Dish Pathfinder Array (TDPA) is a radio interferometer designed to test techniques for 21 cm intensity mapping in the post-reionization universe as a means of measuring large-scale cosmic structure. Using 9 nights of observations 
targeting the North Celestial Pole (NCP) field, totaling approximately 107 hours of integration time, we analyze data in the frequency range 700--800~MHz (corresponding to redshift $z \sim 0.9$). We do the data format conversion, radio frequency interference (RFI) flagging, calibration, imaging and point source subtraction, and foreground removal via Singular Value Decomposition (SVD). 
The spherically averaged power spectrum $\Delta^2(k)$ is obtained. This work successfully establishes and validates a comprehensive data analysis framework for the TDPA. We identify key improvements including sky model refinement, increased integration time, and pipeline optimization that will enable future detection of the 21\,cm signal through auto-correlation and cross-correlation with optical galaxy surveys.
}


   \titlerunning{21 cm Power Spectrum Analysis of NCP Observations with the TDPA}
   \maketitle
%
%
\section{Introduction}

Understanding the distribution of matter in the large-scale structure of the Universe is fundamental to cosmology. It enables us to test and constrain the standard $\Lambda$CDM cosmological model, investigate the nature of dark energy, and probe the physics of the early Universe.

Neutral hydrogen (HI) at the 21\,cm radio wavelength serves as an excellent tracer of matter distribution. In the post-reionization Universe at redshifts $z \lesssim 6$, the remaining HI is predominantly contained within dark matter halos \citep{2018ApJ...866..135V}, meaning that the 21\,cm radiation emitted by HI acts as a tracer of the large-scale structure.
However, at these redshifts, individual HI-rich galaxies are too faint to be detected with high signal-to-noise ratio by traditional galaxy surveys. Instead, HI intensity mapping (IM) measures the integrated 21\,cm emission from unresolved sources across large sky areas, providing a highly efficient method to survey vast cosmic volumes and access redshifts ranging from the local Universe to the epoch of reionization \citep{2001JApA...22...21B,2004MNRAS.355.1339B,2008MNRAS.383.1195W,2008PhRvL.100i1303C,2023ApJ...954..139L,2025ApJS..279...32Y}.

However, HI intensity mapping faces significant observational challenges. The technique records all radiation entering the telescope receiver, of which only a tiny fraction originates from HI. Foreground contamination from Galactic synchrotron and free-free emission, as well as strong extragalactic point sources, exceeds the cosmological HI signal by 3--5 orders of magnitude \citep{2019MNRAS.483.4411M,2021MNRAS.505.1575S,2022MNRAS.509.4923I,2021MNRAS.506.5075M}. Furthermore, instrumental systematics can distort these foreground signals in complex ways, making their removal particularly challenging \citep{2015PhRvD..91h3514S, 2022MNRAS.509.2048S}. RFI from terrestrial sources presents an additional obstacle to HI measurements \citep{2018MNRAS.479.2024H, 2025MNRAS.536.1035E}.

Nevertheless, several experiments have reported promising preliminary results \citep{2013ApJ...763L..20M, 2018MNRAS.476.3382A, 2025arXiv251119620C, 2023arXiv230111943P,2026arXiv260223055T} , while other experiments are currently ongoing or under construction and are expected to deliver results in the near future \citep{2023ApJ...954..139L, 2025ApJS..279...32Y, 2012arXiv1209.1041B, 2016SPIE.9906E..5XN}.

The Tianlai project is an experiment designed to address these challenges in 21~cm intensity mapping \citep{2012IJMPS..12..256C}. It consists of two pathfinder arrays: the Tianlai cylinder pathfinder array \citep{2020SCPMA..6329862L} and the Tianlai dish pathfinder array (TDPA, \citealt{2021MNRAS.506.3455W}). The present work is devoted to the data analysis of the latter. The dish array is equipped with dual linear-polarization feeds and arranged in a compact, centrally condensed configuration \citep{Zhang2016}. The beam profile of the dish antenna has been measured with a drone \cite{UAV2021}, and the cross-couplings between the antennas was studied by simulation \citep{Kwak2024}. The overall performance of the dish array was studied in \citep{2021MNRAS.506.3455W}. 

The Tianlai dishes are steerable, but to mitigate possible variations in ground pick up and other environmental factors, drift scan is the favored observation mode. The North Celestial Pole (NCP) is a point of special interest for drift scan observation, as it does not move, and allows accumulation of large integration time for deep observation. A survey of the NCP region is of particular interest \citep{Perdereau2022}, and complementary to the radio survey, the optical Tianlai-WIYN survey of the NCP region is on-going \citep{Ansari2025}.

In this work, we present results from 9 nights of observations with the Tianlai Dish Pathfinder Array (TDPA), totaling approximately 107 hours of integration time on the center of the NCP field. This is only a small part of the designed NCP survey in \citet{Perdereau2022}, but it may serve as a trial.
We describe our complete data processing pipeline and perform power spectrum estimation to assess the current performance and systematic error budget of the instrument. This paper is organized as follows: Section~\ref{sect:Obs} describes the observations and data characteristics; Section~\ref{sect:data} details the pre-processing, calibration, and imaging procedures; Section~\ref{sect:analysis} presents the foreground-cleaning, temperature-conversion, and power-spectrum analysis; and Section~\ref{sect:discussion} discusses the implications and future improvements.

\section{Observations}
\label{sect:Obs}

\begin{figure}
   \centering
   \includegraphics[width=0.5\linewidth, angle=0]{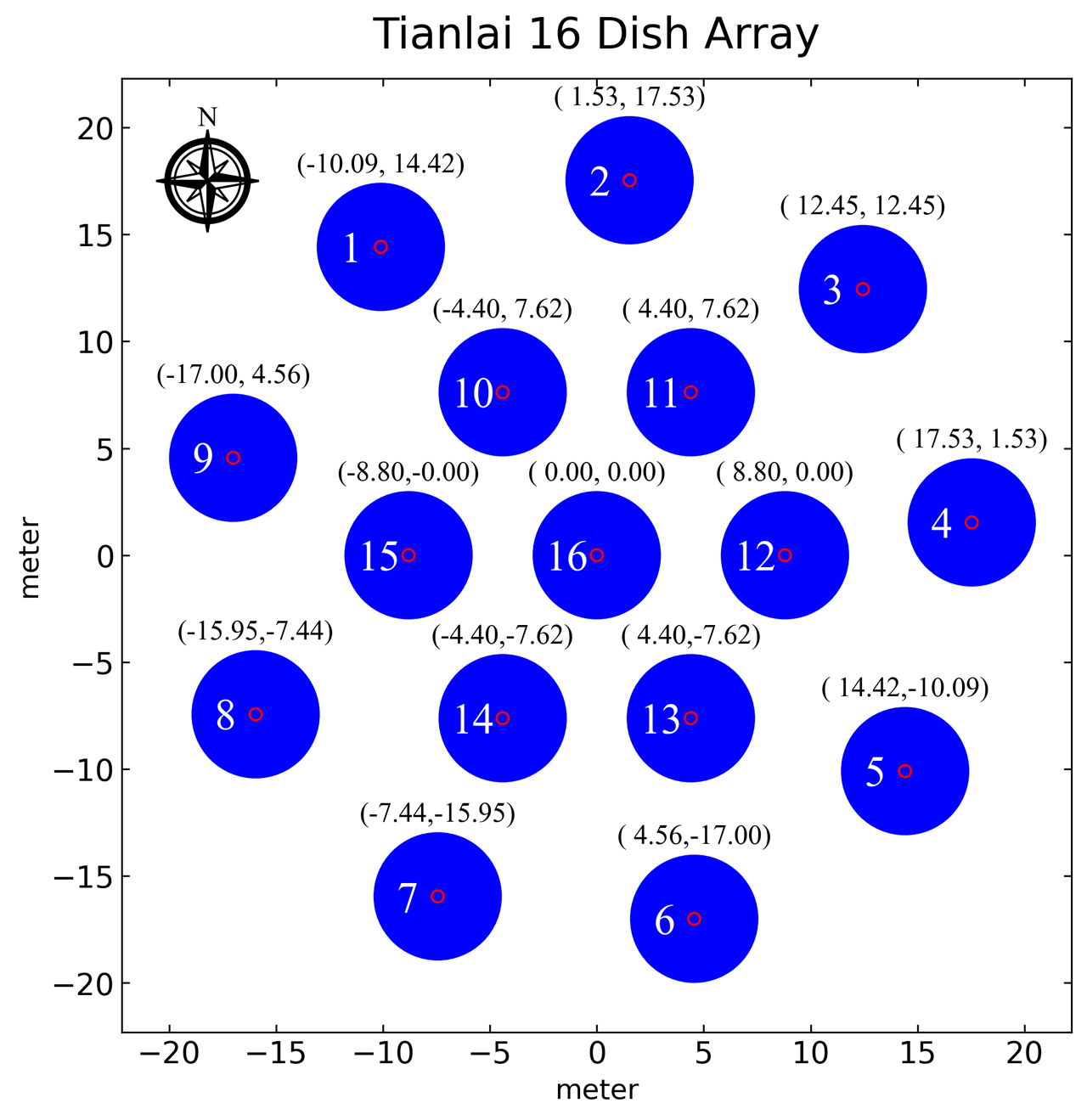}
   \caption{Tianlai Dish Pathfinder Array layout with antenna number and ground coordinates with respect to a reference point (unit:meter).}
   \label{fig:dish_layout}
   \end{figure}

The TDPA is located at a radio quiet site ($44^\circ9'$N, $91^\circ48'$E) in Hongliuxia, Balikun County, Xinjiang Autonomous Region, in northwest China. The array features 16 on-axis dishes, each 6 meters in diameter. As shown in Fig. \ref{fig:dish_layout}, the antennas are arranged in a compact circular configuration, comprising one central dish surrounded by two concentric rings\citep{2022RAA....22l5007Y}.

It operates as a transit interferometer in drift-scan mode, with the dishes pointed at a fixed direction in the meridian plane while the sky drifts through the primary beam.
For a northern-hemisphere transit array, the North Celestial Pole (NCP) is the only field that can be monitored continuously, so long NCP drift scans maximize the signal-to-noise ratio for a given survey time \citep{2021MNRAS.506.3455W}.
Consequently, the array has accumulated extensive drift-scan data on this field. 
Moreover, simulations of planned low-redshift surveys with the TDPA show that deep NCP observations achieve lower noise levels and are weakly affected by mode mixing compared to mid-latitude scans \citep{Perdereau2022}.  
In this work, we analyzed 9 nights of observations from the TDPA, amounting to approximately 107 hours of observation. The data were recorded over a frequency range of 700 to 800 MHz, with a spectral resolution of 0.244 MHz and a temporal resolution of 1 second.
A summary of all observations used for this analysis is provided in Table~\ref{Tab1}.

\begin{table}
\begin{center}
\caption[]{List of all observation nights analyzed in this work. The datasets for 0307 and 0309 correspond to two consecutive days, each lasting approximately 21 hours. Daytime data were removed early in the analysis, and only the analyzed data lengths are shown.}\label{Tab1}


 \begin{tabular}{clcr}
  \hline\noalign{\smallskip}
Data Set & Start time (UTC+0h) & Start time (UTC+8h) & Length (hours)                    \\
  \hline\noalign{\smallskip}
NP\underline{ }20200303  & 2020-03-03 13:38:36 & 2020-03-03 21:38:36     & 10.0  \\ 
NP\underline{ }20200304  & 2020-03-04 13:38:56     & 2020-03-04 21:38:56       & 11.0                  \\
NP\underline{ }20200306  & 2020-03-05 18:19:56     & 2020-03-06 02:19:56     & 6.0                  \\
NP\underline{ }20200307  & 2020-03-07 14:59:01     & 2020-03-07 22:59:01      & 9.0                  \\
 & 2020-03-08 11:59:01     & 2020-03-08 19:59:01      & 12.0 \\
NP\underline{ }20200309  & 2020-03-09 13:21:07     & 2020-03-09 21:21:07     & 10.0                  \\
 & 2020-03-10 12:21:07     & 2020-03-10 20:21:07      & 11.0 \\
NP\underline{ }20200311  & 2020-03-11 14:09:48     & 2020-03-11 22:09:48     & 9.0                  \\
NP\underline{ }20200312  & 2020-03-12 16:02:27    &  2020-03-13 00:02:27     & 8.0                  \\
NP\underline{ }20200313  & 2020-03-13 14:02:08     & 2020-03-13 22:02:08     & 10.0                  \\
NP\underline{ }20200314  & 2020-03-14 13:12:40     & 2020-03-14 21:12:40     & 11.0                  \\
  \noalign{\smallskip}\hline
\end{tabular}
\end{center}
\end{table}

\section{Data reduction}
\label{sect:data}

\begin{figure}
   \centering
   \includegraphics[width=0.8\linewidth, angle=0]{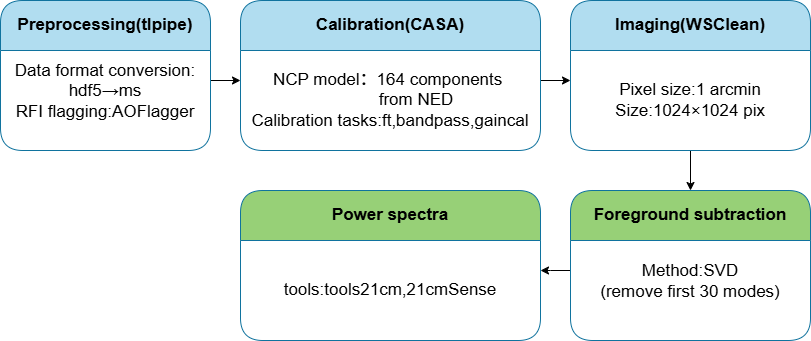}
   \caption{Schematic diagram of the TDPA processing pipeline. Blue blocks indicate the data-processing stages (implemented mainly with tlpipe and standard radio-astronomy packages; Section \ref{sect:data}), while green blocks indicate the analysis stages for foreground subtraction and power-spectrum estimation (implemented mainly with Python packages; Section \ref{sect:analysis}).}
   \label{fig:pipeline}
   \end{figure}

Standard 21 cm intensity-mapping analyses typically proceed through pre-processing, calibration, imaging, foreground subtraction, and power-spectrum estimation (e.g., \citep{2017ApJ...838...65P,2020MNRAS.493.1662M}).
Our pipeline for the TDPA data is summarized in Fig. \ref{fig:pipeline} and comprises five stages: (1) Pre-processing, where format conversion and RFI flagging are performed; (2) Calibration, including self-calibration using sky models; (3) Imaging and point source subtraction; (4) Residual foreground subtraction; (5) Power-spectrum estimation. 

\subsection{Pre-processing}

\begin{figure}
   \centering
   \includegraphics[width=0.7\linewidth, angle=0]{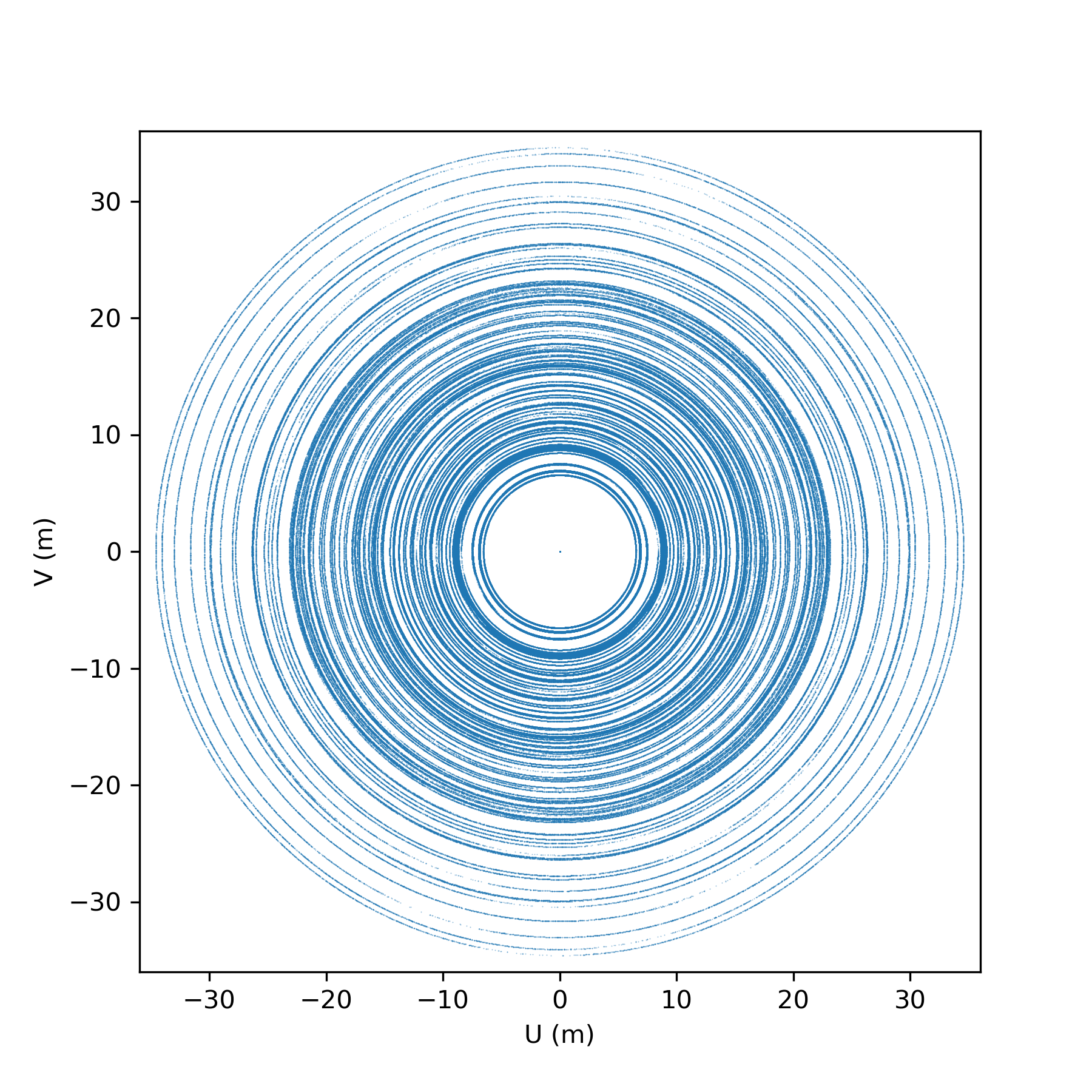}
   \caption{$uv$ coverage constructed from the full TDPA dataset after data format conversion from HDF5 to CASA Measurement Set. The plot shows the distribution of baseline sampling in the $uv$ plane, where each sampled point represents the $(u,v)$ coordinate of a baseline at a given time stamp.}
   \label{fig:uv_coverage}
   \end{figure}

In this work we build our pipeline using the standard radio astronomical software like CASA \citep{PASP.134.114501.} and WSClean \citep{2014MNRAS.444..606O}, which are well tested and easy to use in the calibration and imaging tasks for a small sky region such as the NCP, complementary to the home made data processing pipeline software called tlpipe \citep{2021A&C....3400439Z}.
The data is converted from the original HDF5 format to CASA Measurement Set files to be processed with these software tools. 
Fig.~\ref{fig:uv_coverage} shows the $uv$ coverage constructed from the full dataset after data format conversion. Each sampled point represents the $(u,v)$ coordinate of a baseline at a given time stamp, thereby illustrating the overall sampling of the interferometric measurements in the Fourier plane.

\begin{figure}
   \centering
   \includegraphics[width=15cm, angle=0]{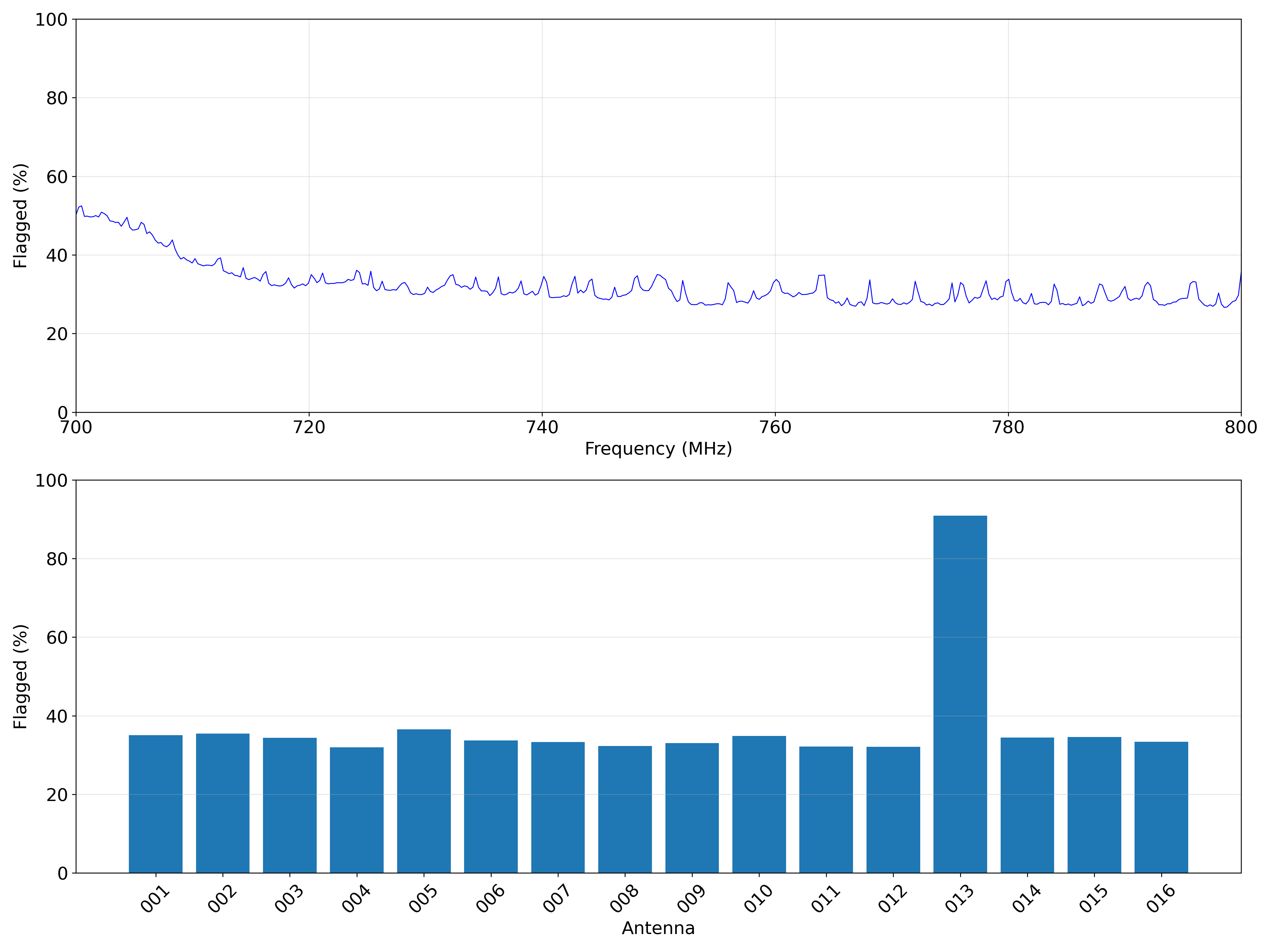}
   \caption{Percentage of flagged data as a function of frequency channel and antenna. 
   Edge channels show higher flagging rates due to stronger system effects. Antenna 013 exhibits consistently high flagging across multiple observations, indicating potential malfunction or damage, resulting in all its data being flagged.}
   \label{Fig4}
   \end{figure}

Subsequently, we performed RFI flagging using AOFlagger \citep{2012A&A...539A..95O}. Fig. \ref{Fig4} displays the data quality flags for the full dataset, including the intervals when the calibration noise source (CNS), which is used to provide relative calibration in the absence of sufficiently bright point sources, was periodically injected. 
Edge channels are more susceptible to systematic effects, resulting in a higher percentage of flagged data. 
Meanwhile, Antenna 013 appears to exhibit errors across multiple observations, suggesting it may be malfunctioning or damaged, and thus all its data have been flagged. 

\subsection{Calibration}
\begin{figure}
   \centering
   \includegraphics[width=0.95\linewidth, angle=0]{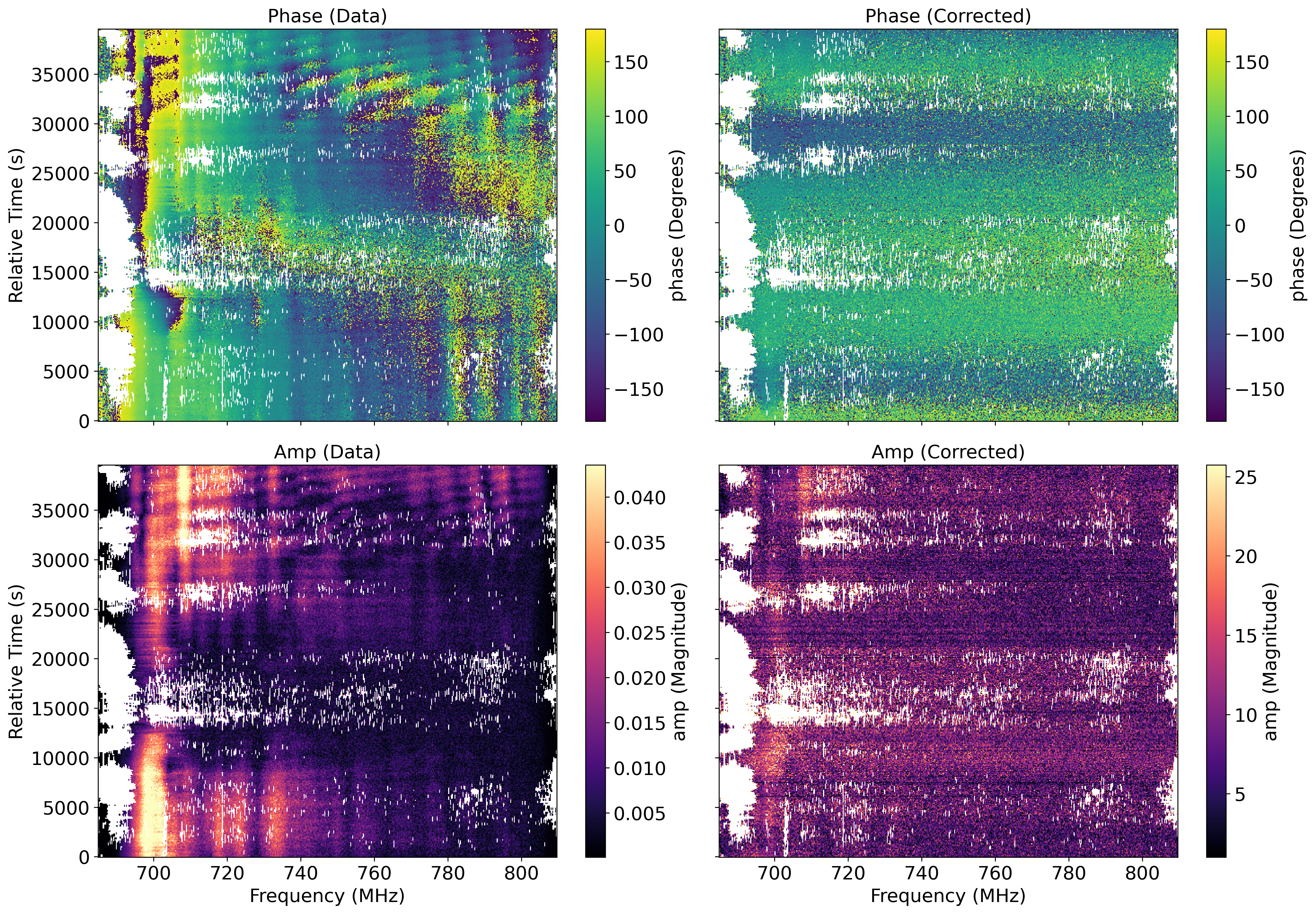}
   \caption{Amplitude and phase of a representative baseline as a function of time and frequency, before (left) and after (right) self-calibration. The calibrated visibilities show significantly reduced phase fluctuations and amplitude variations across the bandpass, demonstrating improved stability after correction for instrumental and propagation effects. 
   }
   \label{fig:self_cal}
   \end{figure}

The calibration begins with a self-calibration \citep{1981MNRAS.196.1067C,1984ARA&A..22...97P} and the main goal is to address instrumental and propagation effects. We first perform gain calibration on the visibilities with a sky model consisting of 164 bright point sources from the NCP sky model. The calibration is carried out by \texttt{ft}, \texttt{bandpass} and \texttt{gaincal} tasks in CASA, where \texttt{ft} inserts the source model into the visibility data, \texttt{bandpass} solves for the frequency-dependent complex bandpass response, and \texttt{gaincal} derives the time-dependent complex antenna gains. In Fig. \ref{fig:self_cal}, we present the amplitudes and phases of a representative baseline as a function of time and frequency, derived from the dataset recorded on March 4, 2020 (NP\underline{ }20200304). Prior to calibration, the raw visibilities typically exhibit significant temporal drifts and frequency-dependent ripples caused by the uncorrected instrumental response and environmental fluctuations. After applying the self-calibration, Fig. \ref{fig:self_cal} demonstrates a marked improvement in stability: the phase fluctuations are largely flattened, and the amplitude variations across the bandpass are significantly suppressed. 

\subsection{Imaging}
\begin{figure}
   \centering
   \includegraphics[width=15cm, angle=0]{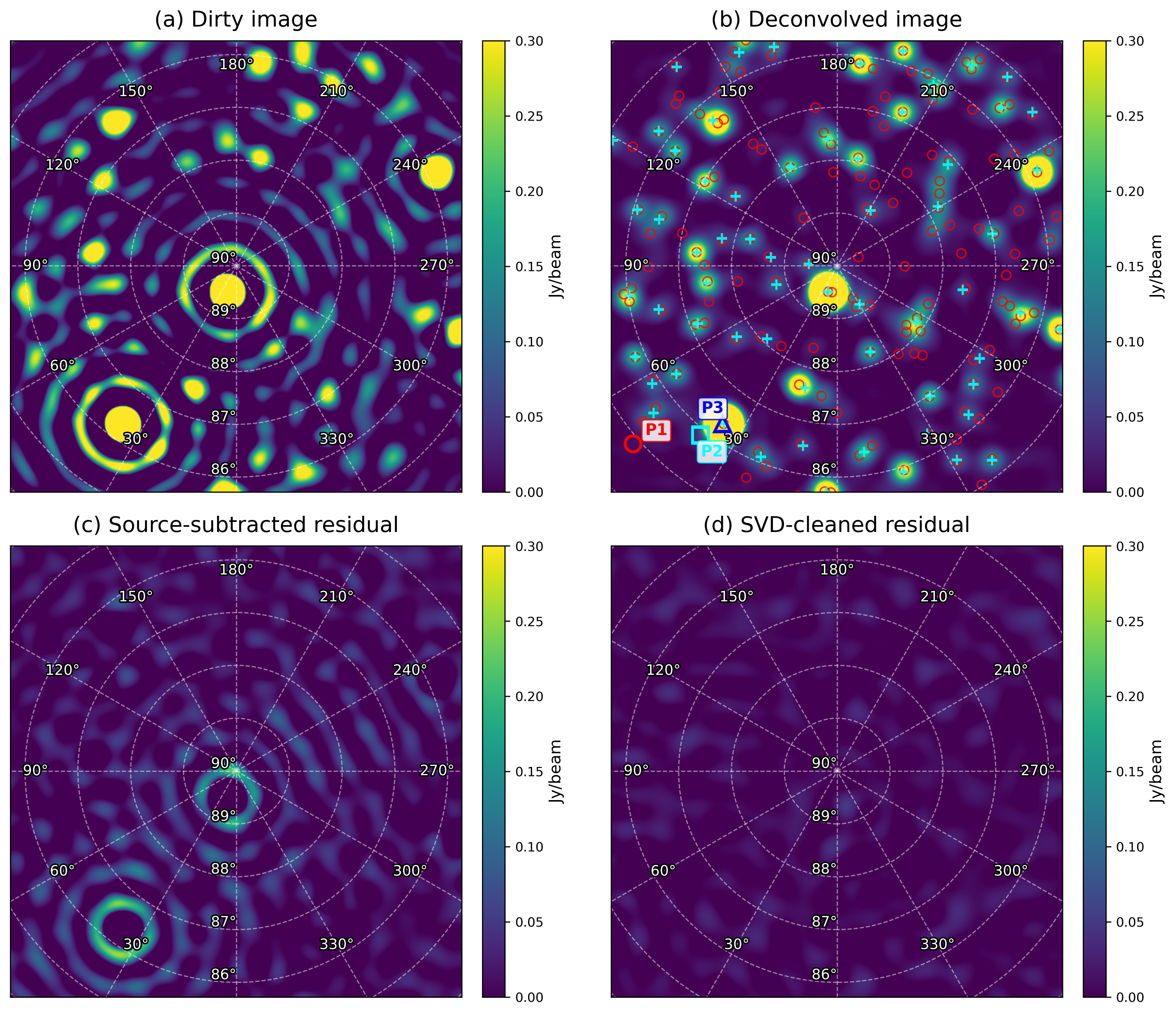}
   \caption{Stokes~I continuum images of the NCP field over 700--800~MHz. 
   \textbf{(a)} Dirty image. 
   \textbf{(b)} Deconvolved image generated using WSClean. Red circles denote the 164 primary calibrators, while cyan crosses mark the sources extracted by PyBDSF. The three selected pixels, P1, P2, and P3, are highlighted, and their frequency spectra are presented in the subsequent panels. 
   \textbf{(c)} Residual map after source subtraction. 
   \textbf{(d)} Residual map after source subtraction and SVD-based foreground removal, with the first 30 SVD modes removed.}
   \label{fig:ncp_map}
\end{figure}

The calibrated visibilities are transformed into images and deconvolved using the multiscale CLEAN algorithm implemented in WSClean. The imaging is performed on the $700$--$800\,\mathrm{MHz}$ band of our dataset, and we use multi-frequency synthesis to combine the frequency channels into a single continuum map. To ensure that the synthesized beam is adequately sampled, we set the pixel size to $1\text{ arcmin}$, providing sufficient resolution relative to the array's longest baselines. The initial image dimensions are configured to $1024 \times 1024$ pixels, resulting in a wide field of view (FoV) of approximately $17^\circ \times 17^\circ$. 
However, retaining such a large FoV can introduce significant errors, as regions extending far beyond the main beam suffer from severe signal attenuation and low signal-to-noise ratios (SNR). In the 700–800 MHz band, the main beam's full width at half maximum (FWHM) fluctuates between 3.8° and 4.6°, averaging around 4.2° \citep{2021MNRAS.506.3455W}). To optimize the signal-to-noise ratio while minimizing the risk of excluding valid emission, we restrict our analysis to the central 512 $\times$ 512 pixel region, corresponding to an $8.5^\circ \times 8.5^\circ$ FoV, as illustrated in Fig.~\ref{fig:ncp_map}. This region is approximately twice the beam FWHM and therefore comfortably encompasses the reliable main beam response. Fig.~\ref{fig:ncp_map} shows the imaging results at successive processing stages: the dirty image in panel (a), the deconvolved image generated with WSClean in panel (b), the residual map after source subtraction in panel (c), and the residual map after subsequent SVD cleaning in panel (d). All subsequent radio-source identification and categorization are performed within this optimized region. Specifically, the red circles in panel (b) mark the positions of the 164 bright point sources in the initial sky model used for calibration. To facilitate successive foreground removal, we further employed PyBDSF \citep{2015ascl.soft02007M} for automated source extraction; the newly identified sources are indicated by cyan crosses.
We evaluated the astrometric accuracy by cross-matching the PyBDSF catalog with the reference model. Our analysis reveals that the positional offsets for the majority of sources are contained within $10\text{ arcmin}$. Furthermore, the distribution of these offsets appears random without significant systematic drift, thereby confirming the robustness of our global phase calibration.

Gain-corrected sky-model visibilities are computed after calibration by applying the gain solutions to the predicted sky-model visibilities and subsequently subtracting them from the observed visibilities. Panel (c) of Fig.~\ref{fig:ncp_map} shows the residual image of the NCP field after source subtraction. While most sources are removed successfully, residual emission remains around the brightest sources, with flux densities ranging from $-1.2$ to $+0.2$~Jy/beam.

%

\section{Foreground subtraction and power spectrum estimation}
\label{sect:analysis}

Although the calibration and point-source subtraction described in Section~\ref{sect:data} can effectively remove the discrete sky components, the residual visibilities are still dominated by diffuse foreground emission. 
At the frequencies relevant to this work, the leading contaminants are Galactic synchrotron radiation and free--free emission, together with the integrated contribution from unresolved extragalactic sources. 
These foregrounds are typically brighter than the expected cosmological 21\,cm signal by about 4--5 orders of magnitude \citep{2005ApJ...625..575S, 2006PhR...433..181F}. 

A crucial distinction, however, lies in their spectral behavior: astrophysical foregrounds are generally smooth functions of frequency, while the 21\,cm signal varies rapidly along the line-of-sight direction due to the underlying density and ionization fluctuations. 
This contrast motivates a broad class of foreground-mitigation techniques that aim to identify and remove spectrally smooth components, while preserving the spectrally fluctuating cosmological signal \citep{2020PASP..132062001L}.

Among the various approaches, blind component-separation methods are widely adopted in 21\,cm analyses. 
Their main appeal is that they do not rely on an accurate parametric model of the foreground emission, nor do they require a detailed description of how the instrument and calibration residuals imprint frequency-dependent systematics onto the data. 
Given that our knowledge of low-frequency sky emission and its coupling to instrumental chromaticity remains incomplete, and that the intensity mapping/EoR analysis pipeline is still actively evolving, it is both practical and robust to prioritize such blind methods \citep{2017MNRAS.464.4938W,2018MNRAS.476.3382A}.

In this section, we first apply a Singular Value Decomposition (SVD) \citep{2014ShlensPCA} based technique to identify and remove the dominant foreground modes in the residual image. 
We then convert the cleaned image into brightness temperature units, and finally estimate the 21\,cm power spectrum from the foreground-subtracted data.

\subsection{SVD foreground cleaning on the map cube}
\label{subsect:svd_clean}

We apply an SVD-based blind foreground cleaning directly to the image-domain data cube. 
The calibrated and imaged cube has an angular size of $N_{\rm RA}\times N_{\rm Dec}$ pixels for each frequency channel, where $N_{\rm RA}$ and $N_{\rm Dec}$ denote the numbers of pixels along the right ascension (RA) and declination (Dec) directions, respectively. In our case, $N_{\rm RA}=N_{\rm Dec}=512$, i.e. each map has $512\times512$ angular pixels, and $N_{\nu}=412$, corresponding to 700--800 MHz.

For each retained frequency channel, the 2D sky map is flattened into a one-dimensional vector, and the full cube is packed into a matrix
\begin{equation}
\mathbf{M}\in \mathbb{R}^{N_{\nu}\times N_{\theta}},
\qquad
N_{\nu}=412,
\qquad
N_{\theta}=N_{\rm RA}N_{\rm Dec}=512\times512.
\end{equation}

We compute the singular value decomposition
\begin{equation}
\mathbf{M}=\mathbf{U}\mathbf{\Sigma}\mathbf{V}^{T},
\end{equation}
where $\mathbf{U}\in\mathbb{R}^{N_{\nu}\times N_{\nu}}$ and $\mathbf{V}\in\mathbb{R}^{N_{\theta}\times N_{\nu}}$ contain the left and right singular vectors and $\mathbf{\Sigma}$ is diagonal with singular values sorted in descending order.
The leading modes capture the largest-variance spectral structures, which are dominated by spectrally smooth foreground emission (and any residual smooth instrumental systematics), while the 21\,cm signal and thermal noise are distributed across a much larger number of modes.

Foreground cleaning is implemented by removing the first $N_{\rm fg}$ modes, i.e. setting the largest $N_{\rm fg}$ singular values to zero and reconstructing the residual:
\begin{equation}
\mathbf{\Sigma}'_{ii}=
\begin{cases}
0, & i\le N_{\rm fg},\\
\mathbf{\Sigma}_{ii}, & i> N_{\rm fg},
\end{cases}
\qquad
\mathbf{M}_{\rm clean}=\mathbf{U}\mathbf{\Sigma}'\mathbf{V}^{T}.
\end{equation}
The cleaned matrix $\mathbf{M}_{\rm clean}$ is then reshaped back into a three-dimensional cube for the subsequent power spectrum estimation.

\begin{figure}
   \centering
   \includegraphics[width=15cm, angle=0]{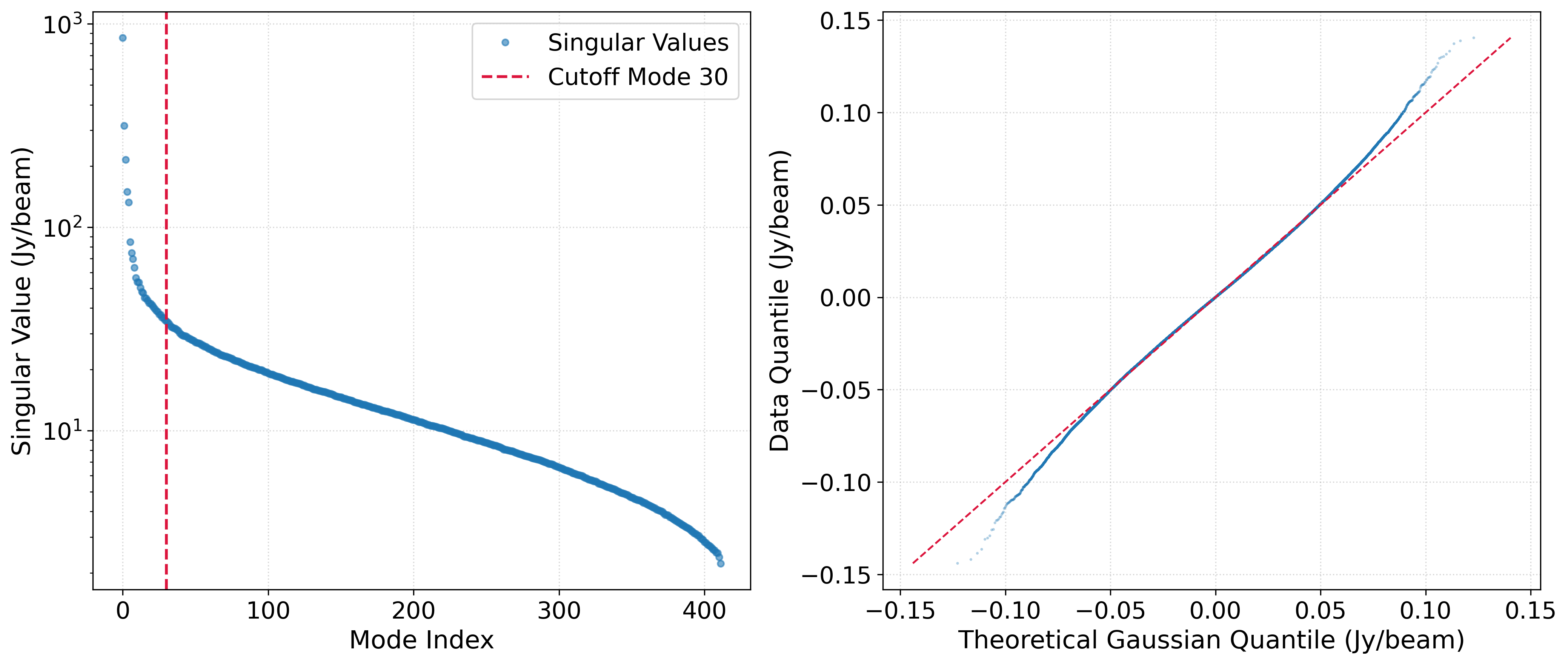}
   \caption{Left: singular-value spectrum of the SVD foreground cleaning. The singular values decrease steeply at low mode number and begin to flatten around $N_{\rm fg}=30$, indicating a transition from dominant foreground modes to a more slowly varying tail. Right: quantile--quantile (Q--Q) plot of the residual voxel values after SVD cleaning with $N_{\rm fg}=30$, compared with a Gaussian reference distribution. The central part is approximately linear, indicating that the bulk of the residuals is close to Gaussian, while mild deviations remain in the tails.}
   \label{fig:svd_qq}
   \end{figure}

In our analysis, we find that small values of $N_{\rm fg}$ leave substantial residual foreground leakage, whereas excessively large values of $N_{\rm fg}$ remove an increasing fraction of the cosmological signal and consequently enlarge the final uncertainties \citep{2026CunningtonRevealing}. We therefore select $N_{\rm fg}=30$ as a conservative compromise. This choice is guided primarily by the singular-value spectrum shown in Fig.~\ref{fig:svd_qq}. The left panel exhibits a clear transition around $N_{\rm fg}\sim 30$, from a steep, foreground-dominated regime to a more slowly varying tail, suggesting that the leading smooth foreground modes are largely captured within the first $\sim 30$ components. This interpretation is further examined using the Q--Q plot, the residual map, and the representative residual spectra discussed below.

The right panel of Fig.~\ref{fig:svd_qq} shows a quantile--quantile (Q--Q) plot \citep{1968WilkProbability}, which provides a simple diagnostic of how closely the residual voxel distribution follows a Gaussian form. To construct this plot, we collect all finite voxel values in the SVD-cleaned residual cube, flatten them into a one-dimensional sample $\{x_j\}_{j=1}^{N}$, and sort them as $x_{(1)} \le x_{(2)} \le \cdots \le x_{(N)}$. We then compute the sample mean $\hat{\mu}$ and standard deviation $\hat{\sigma}$, and compare the ordered residuals $x_{(i)}$ with the corresponding theoretical Gaussian quantiles
\[
q_i=\hat{\mu}+\hat{\sigma}\,\Phi^{-1}(p_i),\qquad p_i=\frac{i-0.5}{N},
\]
where $\Phi^{-1}$ is the inverse cumulative distribution function of the standard normal distribution. If the residuals are Gaussian-distributed, the points $(q_i, x_{(i)})$ should lie close to the one-to-one line $y=x$. Thus, the approximate linearity seen in the central part indicates that the bulk of the cleaned residuals is close to Gaussian, while the remaining deviations in the tails suggest residual non-Gaussian contamination, such as unsubtracted foreground structures, calibration artifacts, or imaging residuals. This behavior is consistent with the removal of the dominant foreground structures. As illustrated in panel (d) of Fig.~\ref{fig:ncp_map}, the residual map after SVD cleaning shows that the residuals associated with the brightest sources, which are still prominent in panel (c), have been effectively eliminated.

\begin{figure}
   \centering
   \includegraphics[width=15cm, angle=0]{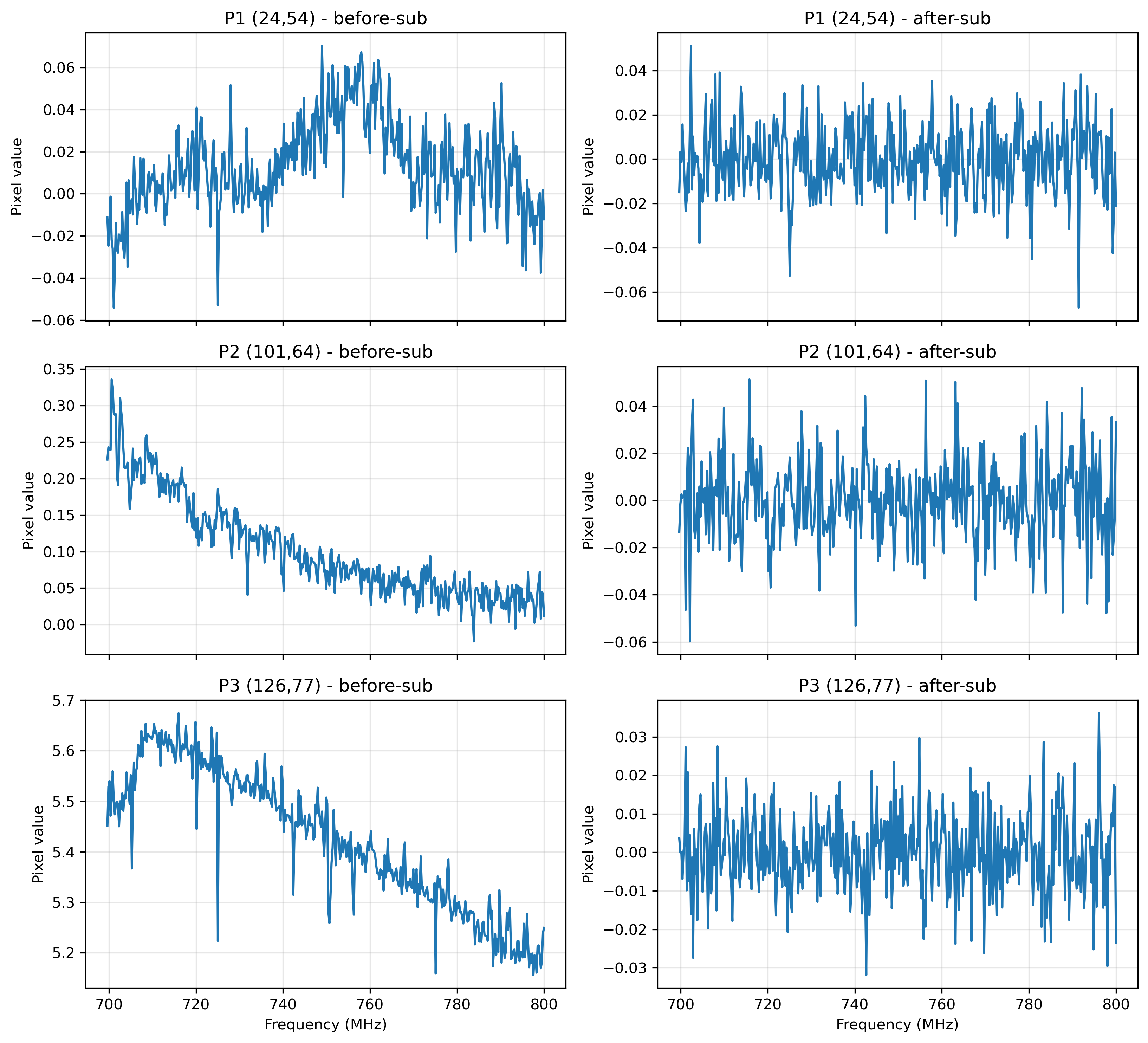}
   \caption{Spectra extracted from the three representative pixels marked in the panel (b) of Fig.~\ref{fig:ncp_map}: P1 corresponds to a source-free region, P2 to the edge of a source, and P3 to the center of a bright source. For each pixel, the left column shows the spectrum before foreground removal, while the right column shows the corresponding residual spectrum after SVD cleaning with $N_{\rm fg}=30$. The strong suppression of the smooth spectral structures demonstrates the effectiveness of the foreground subtraction.}
   \label{fig:spectra}
   \end{figure}

This conclusion is also supported by the spectra extracted from the three representative pixels, P1, P2, and P3, marked in panel (b) of Fig.~\ref{fig:ncp_map} and shown in Fig.~\ref{fig:spectra}. These three pixels correspond respectively to a source-free region, the edge of a source, and the center of a bright source. Before foreground removal, their spectra exhibit prominent frequency-dependent structures. For the pixels located at the source center and source edge, these features mainly reflect the spectral behavior of the bright point source itself, together with source-associated extended structures introduced by the synthesized beam and imaging response. In contrast, for the source-free pixel, the observed spectral structure is more likely due to a combination of diffuse foreground contamination, unresolved faint sources, and residual calibration/imaging systematics. After SVD cleaning with $N_{\rm fg}=30$, the spectral structures in all three cases are strongly suppressed, and the residual spectra become flatter and largely noise-like around zero, indicating effective suppression of the dominant foreground- and source-related contaminants.

\subsection{Conversion to Brightness Temperature}
\label{subsect:jybeam_to_k}

The WSClean imager outputs the deconvolved Stokes-$I$ map cube in units of Jy/beam. For intensity mapping and power-spectrum estimation, it is more convenient to work in terms of brightness temperature, since the cosmological 21\,cm signal is naturally described as fluctuations in $T_{\rm b}$ (K or mK). We therefore convert the map cube from Jy/beam to Kelvin using the Rayleigh--Jeans relation.

A pixel value in Jy/beam represents the flux density integrated over the clean beam. 
To convert this quantity to specific intensity $I_\nu$ (Jy\,sr$^{-1}$), we divide by the beam solid angle $\Omega_{\rm beam}$. Approximating the clean beam as an elliptical Gaussian with major and minor full-width at half-maximum (FWHM) $\theta_{\rm maj}$ and $\theta_{\rm min}$, the beam solid angle is
\begin{equation}
\Omega_{\rm beam}(\nu) = \frac{\pi}{4\ln 2}\,\theta_{\rm maj}(\nu)\,\theta_{\rm min}(\nu),
\end{equation}
where $\theta_{\rm maj}$ and $\theta_{\rm min}$ are in radians. The corresponding specific intensity is then
\begin{equation}
I_\nu = \frac{S_\nu}{\Omega_{\rm beam}(\nu)} \quad [{\rm Jy\,sr^{-1}}],
\end{equation}
with $S_\nu$ the map value in Jy/beam.

Using the Rayleigh--Jeans approximation, the brightness temperature is related to specific intensity by
\begin{equation}
T_{\rm b}(\nu) = \frac{c^2}{2k_{\rm B}\nu^2}\,I_\nu,
\end{equation}
where $c$ is the speed of light, $k_{\rm B}$ is the Boltzmann constant, and $\nu$ is the observing frequency. Converting Jy to SI units via $1~{\rm Jy}=10^{-26}~{\rm W\,m^{-2}\,Hz^{-1}}$, the full conversion applied to each frequency channel can be written as
\begin{equation}
T_{\rm b}(\nu) =
\left(\frac{S_\nu}{\Omega_{\rm beam}(\nu)}\right)\times 10^{-26}\,
\frac{c^2}{2k_{\rm B}\nu^2}.
\end{equation}
After this step, the map cube is expressed in Kelvin, providing the appropriate physical units for subsequent analysis and power spectrum estimation.

\subsection{Power Spectrum Estimation}
\label{subsect:ps_estimation}

We estimate the 21 cm power spectrum from the brightness-temperature cube in Fourier space. The three-dimensional power spectrum is defined as a function of the wavevector $\mathbf{k}$ by
\begin{equation}
P(\mathbf{k}) \equiv \frac{V_{\rm vox}^2}{V_{\rm box}}\,\big|\widetilde{T}(\mathbf{k})\big|^2,
\label{eq:pk_def}
\end{equation}
where $V_{\rm box}=L_xL_yL_z$ is the total comoving volume of the cube, and $V_{\rm vox}=V_{\rm box}/N_{\rm vox}$ is the comoving volume of a single voxel \citep{2020JOSS....5.2363G}. Here $N_{\rm vox}=N_xN_yN_z$ is the total number of voxels, with $N_x$ and $N_y$ the transverse pixel numbers and $N_z$ the number of frequency channels used in the analysis.
The Fourier transform $\widetilde{T}(\mathbf{k})$ is computed from the discretised field $T(\mathbf{r}_n)$ on the grid as
\begin{equation}
\widetilde{T}(\mathbf{k}) \equiv \sum_{n=1}^{N_{\rm vox}} T(\mathbf{r}_n)\,e^{-i\mathbf{k}\cdot \mathbf{r}_n},
\label{eq:tfft_def}
\end{equation}
where $\mathbf{r}_n$ denotes the comoving coordinate of voxel $n$.
With the box dimensions expressed in ${\rm Mpc}/h$, Eq.~(\ref{eq:pk_def}) yields $P(\mathbf{k})$ in units of ${\rm K}^2({\rm Mpc}/h)^3$.

The Fourier-space coordinates are constructed from the discrete mode indices and the box dimensions as
\begin{equation}
k_x = \frac{2\pi n_x}{L_x},\qquad
k_y = \frac{2\pi n_y}{L_y},\qquad
k_\parallel \equiv |k_z| = \left|\frac{2\pi n_z}{L_z}\right|,
\label{eq:kxyz_def}
\end{equation}
and we define the cylindrical decomposition
\begin{equation}
k_\perp \equiv \sqrt{k_x^2+k_y^2}, \qquad k_\parallel \equiv |k_z|.
\label{eq:kper_kpar_def}
\end{equation}

\begin{figure}
   \centering
   \includegraphics[width=0.95\linewidth, angle=0]{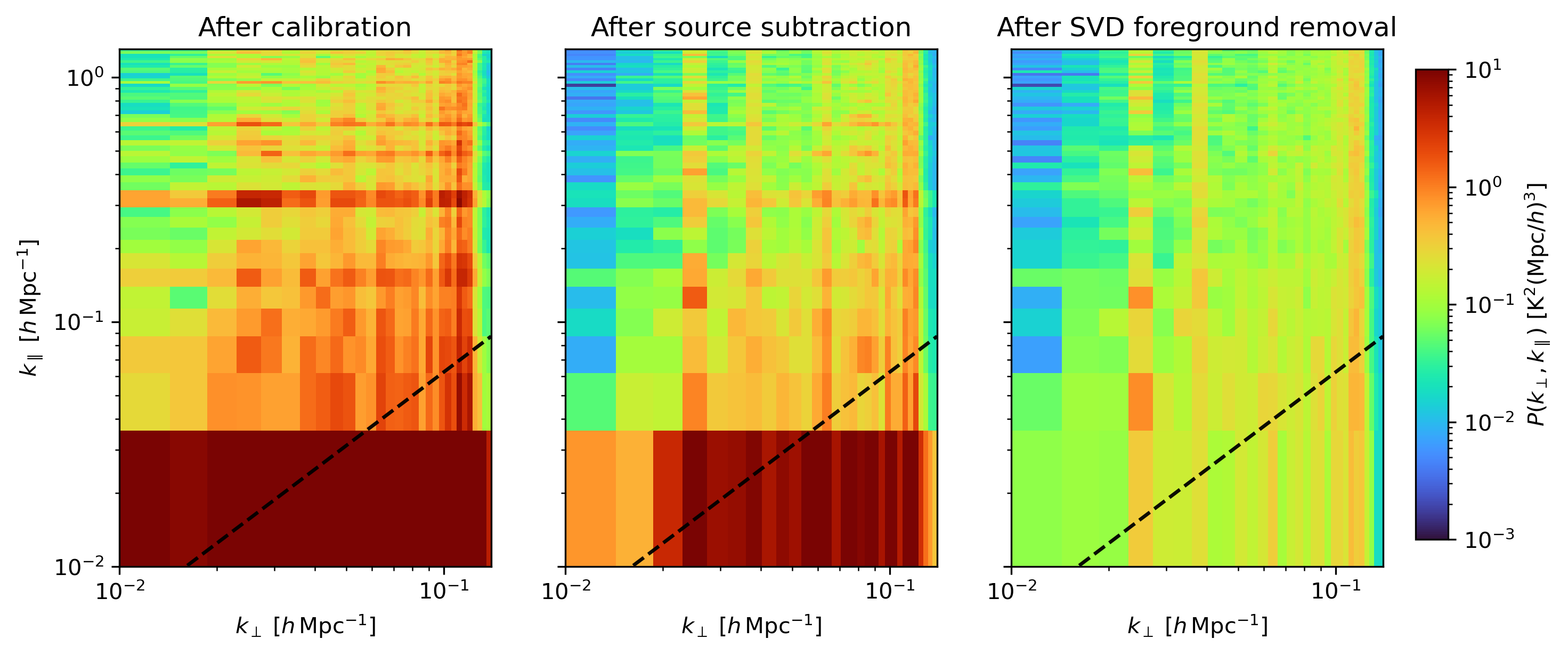}
   \caption{Cylindrical power spectrum $P(k_\perp, k_\parallel)$ at different processing stages. \textit{Left}: After calibration; \textit{Middle}: After point source subtraction; \textit{Right}: After SVD-based foreground cleaning with $N_{\rm fg}=30$ modes removed. The diagonal dashed line indicates the horizon limit ($k_\parallel = 0.56k_\perp$ at $z \approx 0.9$), below which foreground contamination dominates the wedge-shaped region. 
   }
   \label{fig:ps_process}
\end{figure}

We then bin the three-dimensional power into a two-dimensional grid in the $(k_\perp,k_\parallel)$ plane to form the cylindrical power spectrum $P(k_\perp,k_\parallel)$. Fig. \ref{fig:ps_process} presents the cylindrically averaged power spectra of Stokes~$I$ at different steps of the processing. The comparison shows that the excess power is significantly suppressed after source subtraction, while the subsequent SVD foreground cleaning further reduces both the residual foreground contamination and the remaining excess power.

Foreground contamination in interferometric 21\,cm observations is expected to occupy a wedge-shaped region in the cylindrical $(k_\perp, k_\parallel)$ power spectrum, due to instrumental chromaticity that mixes spectrally smooth foreground emission into higher $k_\parallel$ modes. The wedge boundary can be expressed as \citep{2014PhRvD..90b3018L,2010ApJ...724..526D}
\begin{equation}
k_\parallel \;=\; \frac{H_0\,E(z)\,D_M(z)}{c\,(1+z)}\,\sin\theta_0 \; k_\perp ,
\label{eq:horizon_limit}
\end{equation}
where $D_M(z)$ is the transverse comoving distance, $E(z)\equiv H(z)/H_0$ is the dimensionless Hubble expansion function, and $\theta_0$ denotes the maximum angular distance from the phase centre over which foreground emission contributes significantly to the contamination.

In this work, the analysis is restricted to the central $8.5^\circ \times 8.5^\circ$ region of the image cube. Taking half of this field as the characteristic angular extent, we adopt $\theta_0 \simeq 4.2^\circ$, which gives a FoV-limited wedge boundary of $k_\parallel \simeq 0.042\,k_\perp$ at $z\simeq 0.90$. For comparison, in the most extreme case where foreground leakage, including contributions from the side lobes, extends to the entire sky, one has $\sin\theta_0 = 1$ (i.e. $\theta_0 = 90^\circ$), corresponding to the conventional horizon limit $k_\parallel \simeq 0.56\,k_\perp$.

We overplot the horizon limit ($\theta_0=90^\circ$) in the cylindrical power-spectrum figures to indicate the foreground wedge region. For the spherically averaged (1D) power spectrum, we adopt a conservative strategy and compute $P(k)$ using only those $(k_\perp,k_\parallel)$ pixels that lie above the horizon-limit boundary, thereby minimising residual foreground leakage into the final 1D estimate.

Assuming an isotropic signal, we can average $P(k)$ in $k$-bins to create the spherically averaged dimensionless power spectrum defined as:
\begin{equation}
\Delta^2(k) = \frac{k^3}{2\pi^2} \langle P(k) \rangle_k.
\end{equation}

We estimate the noise power spectrum through $\Delta^2_N(k)$ as \citep{2013AJ....145...65P,2012A&A...540A.129A}:
\begin{equation}
\Delta^2_N(k) \approx \frac{k^3}{2\pi^2} X^2 Y \frac{\Omega_{\text{beam}}}{2 t} T_{\text{sys}}^2,
\end{equation}
where $X^2 Y$ is a scalar factor that translates observed units to cosmological distances. $X$ converts angles on the sky to transverse distances, and $Y$ converts bandwidth to line-of-sight distance. Specifically, we have:
\begin{equation}
X = D_A (1 + z) \equiv \int_0^z \frac{c \, dz}{H(z)}, \quad Y = \frac{c(1 + z)^2}{\nu_{21} H(z)}.
\end{equation}
where $\Omega_{\text{beam}}$ is the solid angle of the primary beam of a single element in steradians, $T_{\text{sys}}$ is the system temperature, and $t$ is the total integration time.We estimate the system temperature from our own Cas~A transit observation. Specifically, we use a representative cross-correlation baseline ($1\times6$, Fig. \ref{fig:dish_layout}) and derive the signal-to-noise ratio from the transit profile after averaging five central frequency channels. Using the flux density of Cas~A at 708 MHz, $S_{\rm CasA}=4085$ Jy, we derive an antenna temperature of $T_{\rm A}=29.98$ K. The system temperature is then calculated as
\begin{equation}
T_{\rm sys}=T_{\rm A}\frac{\sqrt{2\Delta\nu t_{\rm int}}}{\mathrm{SNR}},
\end{equation}
where $\Delta\nu = 5\times0.244~\mathrm{MHz}=1.22~\mathrm{MHz}$ , $t_{\rm int}=1~\mathrm{s}$ and SNR = 475.95. This gives a representative value of $T_{\rm sys}=98.39$ K, which is adopted in the following noise power spectrum estimate.

\begin{figure}
   \centering
   \includegraphics[width=0.8\linewidth, angle=0]{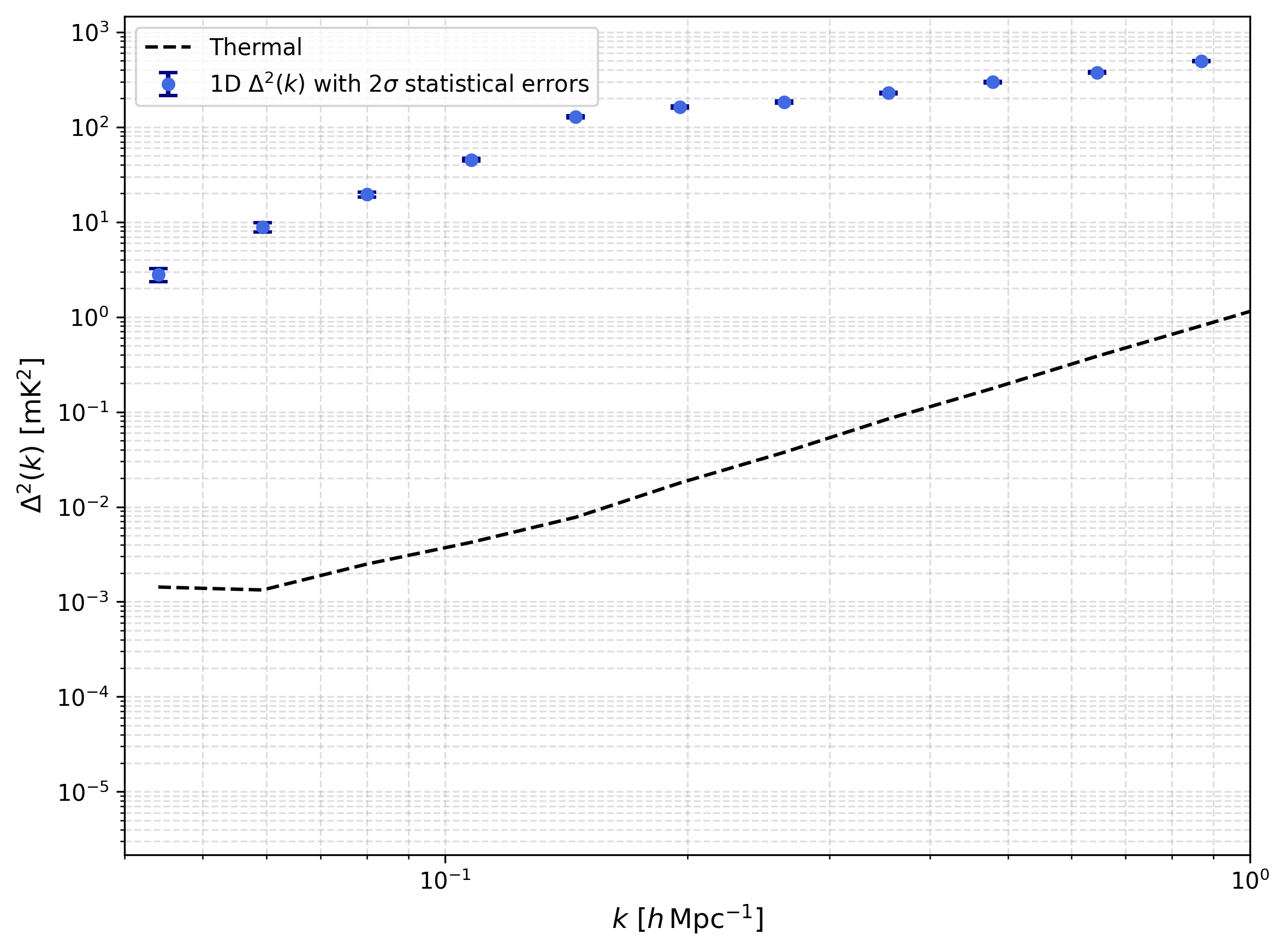}
   \caption{Spherically averaged dimensionless power spectrum $\Delta^2(k)$ after SVD foreground cleaning (blue points with error bars), compared with the theoretical thermal noise power spectrum (dashed line). The error bars denote the $2\sigma$ statistical uncertainty estimated from the standard deviation of the Fourier-mode powers within each $k$ bin. The measured power exceeds the thermal noise expectation by approximately $10^3$--$10^4$ times, indicating significant residual contamination from instrumental systematics, calibration errors, and incomplete foreground removal.}
   
   \label{fig:ps_1d}
\end{figure}

Fig. \ref{fig:ps_1d} presents the spherically averaged dimensionless power spectrum alongside the estimated noise power spectrum. We observe that the measured power exceeds the expected thermal noise level by approximately 3--4 orders of magnitude. Although the SVD cleaning effectively removes the visible foreground wedge from the two-dimensional spectrum, the persistently high power in the one-dimensional spectrum indicates significant foreground leakage into the clean window.

\section{Discussion and Conclusions}
\label{sect:discussion}


In this study, we have presented a comprehensive data analysis pipeline for the Tianlai Dish Pathfinder Array, applying it to nine nights ($\sim 107$~hours) of observations targeting the North Celestial Pole (NCP) field. Our analysis covers the frequency range of 700--800~MHz, corresponding to the post-reionization redshift of $z \sim 0.9$. By employing standard calibration procedures followed by an SVD-based blind foreground subtraction, we successfully obtained deconvolved images and estimated the resulting 21\,cm power spectrum.

Our results demonstrate that the SVD technique, with the removal of $N_{\rm fg}=30$ modes, effectively eliminates the visually prominent foreground wedge in the cylindrical 2D power spectrum $P(k_\perp, k_\parallel)$. However, as shown in Fig. \ref{fig:ps_1d}, the spherically averaged 1D power spectrum $\Delta^2(k)$ reveals a power level approximately $10^3$--$10^4$ times higher than the theoretical thermal noise limit. 

Understanding the origin of this excess power is crucial for the future detection of both the 21\,cm auto-correlation and cross-correlation signals. This residual contamination is likely attributable to several instrumental and modeling non-idealities:
\begin{enumerate}
    \item \textbf{Sky model incompleteness:} Our current calibration relies on a model of only 164 bright sources in the NCP region. This implies that a significant number of fainter point sources remain unsubtracted. Furthermore, bright sources such as Cassiopeia A and Cygnus A can inject power through the antenna sidelobes, contributing to the observed excess.
    \item \textbf{Calibration artifacts:} Imperfect calibration, for example arising from an incomplete or inaccurate sky model, is known to introduce spurious power into the spectrum \citep{2016MNRAS.463.4317P, 2017MNRAS.470.1849E, 2019ApJ...884....1B}. 
    \item \textbf{Instrumental and environmental effects:} Instrumental imperfections, such as bandpass ripples and antenna sidelobe chromaticity, couple smooth foregrounds into high-$k_{\parallel}$ modes. Additionally, low-level radio frequency interference  that escapes detection by AOFlagger can further elevate the noise floor \citep{2019PASP..131k4507W}.
\end{enumerate}

Based on this analysis, we identify several key areas for future improvement: 
(i) \textit{Refining the sky model}: Including fainter compact sources both inside and outside the primary beam will enhance the precision of both calibration and foreground subtraction; 
(ii) \textit{Increasing integration time}: While this work utilizes 107~hours of data, the TDPA has accumulated a much larger dataset. Incorporating the full available data will significantly improve the signal-to-noise ratio; 
(iii) \textit{Pipeline optimization}: We will further optimize AOFlagger settings and CASA calibration parameters (e.g., solution intervals) to achieve superior RFI mitigation and gain stability.
(iv) \textit{Polarization diagnostics}: In future work, we will also form and analyze the Stokes~Q and U map cubes in parallel with Stokes~I. Since polarized foreground emission, after Faraday rotation and instrumental polarization leakage, can introduce frequency-dependent contamination into Stokes~I, this will help assess whether residual polarized foregrounds contribute to the excess power that remains in the measured 1D power spectrum \citep{2010MNRAS.409.1647J,2013ApJ...769..154M}.

Although the current excess noise level prevents direct extraction of the faint cosmological HI signal, this analysis has successfully established and validated a complete data processing pipeline for the TDPA. In the future, by implementing the improvements outlined above and combining our observations with cross-correlation analysis against optical galaxy surveys, we expect to overcome the current systematic error bottleneck. This progress will not only open the scientific discovery window for the Tianlai Array in HI intensity mapping, but will also provide valuable practical experience for future larger-scale radio interferometric arrays designed to probe cosmic evolution. 

\begin{acknowledgements}
We thank the referee for comments and suggestions that improved the paper. This work is supported by the National SKA Program of China (Nos. 2022SKA0110100 and 2022SKA0110101), the National Natural Science Foundation of China (NSFC) International (Regional) Cooperation and Exchange Project (No. 12361141814), the NSFC (No. 12303004,12203061,12473094,12273070.),the NSFC innova-
tion group grant 12421003, the Specialized Research Fund for State Key Laboratory of Radio Astronomy and Technology, and the National Astronomical Observatories, Chinese Academy of Science (No. E5ZB0901). Y.L. acknowledges the support of the National Natural Science Foundation of
China (Nos. 12533001,12473091). This work is also supported by science research grants from the China Manned Space Project with grant Nos. CMS-CSST-2021-B01 and CMS- CSST-2021-A01.

\end{acknowledgements}

\bibliographystyle{raa} 
\bibliography{references}   

\label{lastpage}

\end{document}